\documentclass{aa}%
\usepackage{rotating}
\usepackage{amsmath}
\usepackage{natbib}
\usepackage{graphicx}
\bibpunct{(}{)}{;}{a}{}{,} 

\begin{document}
\title{An Artificial Neural Network Approach to the Solution of Molecular Chemical Equilibrium}

\author{A. Asensio Ramos\inst{1} \and H. Socas-Navarro\inst{2}}
\institute{Istituto Nazionale di Astrofisica (INAF) Osservatorio Astrofisico di Arcetri, Largo Enrico Fermi 5, 50125 Florence, Italy
\and
High Altitude Observatory, NCAR\thanks{The National Center for Atmospheric
  Research (NCAR) is sponsored by the National Science Foundation}, 3450
Mitchell Ln, Boulder CO 80307-3000, USA}

\date{Received <date> / Accepted <date>}

\abstract{A novel approach is presented for the solution of instantaneous chemical equilibrium problems. The chemical equilibrium can be considered, due to its intrinsically local character, as a mapping of the three-dimensional
parameter space spanned by the temperature, hydrogen density and electron density into many one-dimensional spaces representing the
number density of each species. We take advantage of the ability of artificial neural networks to approximate non-linear functions
and construct neural networks for the fast and efficient solution of the chemical equilibrium problem in typical stellar
atmosphere physical conditions. The neural network approach has the advantage
of providing an analytic function, which can be rapidly evaluated. The
networks are trained with a learning set (that covers the entire
parameter space) until a relative error below 1\% is reached. It has been verified that the networks are not overtrained by using an
additional verification set. The networks are then applied to a snapshot of realistic three-dimensional convection simulations
of the solar atmosphere showing good generalization properties.
}

\titlerunning{Neural Network approach to ICE}
\maketitle
\section{Introduction}
Molecules are usually found in highly dynamic systems (e.g., the solar atmosphere, winds of AGB stars, ...) and their formation
is influenced by the time variation of the physical conditions in the
medium. In many situations the dynamical timescales are
much slower than the timescales of molecular formation, and the approximation of Instantaneous Chemical Equilibrium (ICE) can
be used with extremely good results \citep[][ etc.]{russell34,tsu64,tsu73,mccabe79}. Under the ICE approximation, molecules and ions are assumed to form instantaneously and
their abundances depend only upon the local temperature and density. Another consequence of this assumption is that the specific
reaction mechanisms that create and destroy a given molecule are irrelevant and only the molecule and its
constituents are important. Concerning the solar atmosphere, it has been recently shown with the comparison between
ICE calculations and detailed chemical evolution calculations that the formation of the strong CO lines around 4.7 $\mu$m can be well
described using the ICE approximation for heights below $\sim$700 km \citep{asensio03}. On the other hand, chemical evolution effects in
regions of the atmosphere above this height are of importance in setting the CO abundance.

In order to calculate the atomic and molecular number densities using the ICE approximation one needs to solve
a non-linear system of equations (see below). Although efficient numerical methods exist for solving this kind of systems, the computing time
when the ICE problem has to be solved in a large amount of points (e.g.,
dense grids, multi-dimensional geometries, iterative inversions, etc) becomes
prohibitive. It is then very important to develop a
numerical method for the rapid solution of chemical equilibrium problems.

Artificial Neural Networks (ANNs) have proven to be a powerful approach to a
broad variety of problems \citep[see, e.g.,][]{B96}.
In the solar
community, they have been recently applied to the problem of inferring the magnetic field from observations of the polarization
profiles of selected atomic lines \citep{carroll01,socas_navarro03,socas_navarro05}.
In this paper, we make use of the ability of ANNs with one hidden layer to approximate any
non-linear continuous function \citep[e.g.,][]{jones90,blum91} to solve the ICE problem.

The structure of the paper is as follows. Section \ref{sec_ICE} describes the ICE approximation and its standard solution. Section
\ref{sec_ANN} describes our approach to the ICE problem using ANNs, discussing how it can be trained to associate a combination of
physical parameters with the number density of each species included in the problem. Section \ref{sec_comparison} details how the
trained network can be applied to solve the ICE problem in realistic convection simulations of the solar atmosphere. In section
\ref{sec_dependence} we show the dependence of the outputs of the ANN on the
physical parameters, which can be easily done because of the
intrinsic analytic character of the mapping generated by the ANN. Finally, the most relevant conclusions are summarized in Section
\ref{sec_conclusion}.

\section{Instantaneous chemical equilibrium}
\label{sec_ICE}
\subsection{Basic equations}
Consider three elements $A$, $B$ and $C$ that stick together to form a molecule $A_aB_bC_c$, in which $a$, $b$ and $c$
are the stoichiometric coefficients indicating the number of times an element is present in the molecule. Because the explicit
reactions which form the molecule are irrelevant under the ICE
approximation, one needs only to consider the following dissociative process
for the formation of the molecule $A_aB_bC_c$:
\begin{equation}
\label{eq_dissociative_equilibrium}
A_aB_bC_c \Leftrightarrow aA + bB + cC.
\end{equation}
This reaction is characterized by its dissociative reaction constant or equilibrium constant. It is given by the ratio between the
product of the partial pressures of the individual elements and the partial pressure of the molecule:
\begin{equation}
\label{eq_equilibrium_constant}
K_p(T) = \frac{P_A^a P_B^b P_C^c}{P_{ABC}},
\end{equation}
where $K_p(T)$ is a function of the temperature and $P_i$ the partial
pressure of the species $i$.
For example, the equilibrium constants for the dissociation of CO and H$_2$ are
given by
\begin{equation}
\label{eq_equilibrium_constant_CO_H2}
K_p^{CO}(T) = \frac{P_C P_O}{P_{CO}}, \qquad \qquad K_p^{H_2}(T) = \frac{P_H^2}{P_{H_2}}.
\end{equation}
The partial pressure is usually related
to the number density $n_i$ of a given species by the ideal gas equation:
\begin{equation}
\label{eq_ideal_gas}
P_i = n_i k T,
\end{equation}
where $k$ is the Boltzmann constant and $T$ the local temperature. This
partial pressure is the pressure of the gas if no other species were present.
The sum of all the partial
pressures is the total pressure in the medium.

Following the same line of reasoning, one can consider the ionization equilibria for the atomic species:
\begin{align}
\label{eq_ionization_equilibrium}
A & \Leftrightarrow A^+ + e^- \nonumber \\
A + e^- &\Leftrightarrow A^-,
\end{align}
with their corresponding equilibrium constants:
\begin{equation}
\label{eq_ionization_constant}
K_{A^+}(T) = \frac{P_{A^+} P_{e^-}}{P_{A}} \qquad K_{A^-}(T) = \frac{P_{A^-}}{P_A P_{e^-}}.
\end{equation}
We now define the \emph{fictitious pressure} $P(i)$ as the pressure exerted
by element $i$ if all the gas were in the
neutral form of the atomic species $i$. It is customary to specify the partial
pressures of all the atomic species in terms of
the fictitious pressure of hydrogen $P(H)$. The coefficient relating one and the other is the abundance of each element with respect
to hydrogen, $A(i)$, so $P(i)=A(i) P(H)$. The abundance of each atomic species depends on the metallicity of the star we are
considering. For the solar case, we have used the standard solar abundances
given by \citet{grevesse84}.

Under ICE conditions, the number of atoms and molecules are obtained by solving the conservation of mass and
the chemical equilibrium
conditions given by Eq. (\ref{eq_equilibrium_constant}). The conservation of mass establishes that the sum of the partial pressures of
all the species containing a given atomic element (taking into account the stoichiometry) equals the fictitious pressure of the
given element:
\begin{equation}
\label{eq_conservation_mass}
P(i) = P_i + P_{i^+} + P_{i^-} + \sum_k \omega_k^i P_k,
\end{equation}
where $P_i$, $P_{i^+}$ and $P_{i^-}$ are the partial pressure of the neutral element $i$, and the ionized elements $i^+$ and $i^-$,
respectively, while $P_k$ is the partial pressure of molecule $k$ which has species $i$ in its composition. $\omega_k^i$ is
the stoichiometric coefficient which indicates the number of times element $i$ appears in molecule $k$. The sum is extended over
all the molecules in which the element $i$ takes part. The previous equation can be rewritten with the aid of Eqs.
(\ref{eq_equilibrium_constant}) and (\ref{eq_ionization_constant}):
\begin{equation}
\label{eq_conservation_mass_2}
P(i) = P_i + K_{i^+} \frac{P_i}{P_{e^-}} + K_{i^-} P_i P_{e^-} + \sum_k \omega_k^i \frac{P_i^{\omega_k^i} P_j^{\omega_k^j} \cdots
P_l^{\omega_k^l}}{K_p^{(k)} (T)},
\end{equation}
where $i$, $j$, \ldots, $l$ are the atomic species composing molecule $k$, $\omega_k^i+\omega_k^j+\ldots+\omega_k^l=n_k$ is
the number of atoms of molecule $k$ and $K_p^{(k)} (T)$ is its equilibrium
constant. Charged molecules can be easily included
by taking into account their dissociation and the ionization equilibrium,
simultaneously. Consider the dissociation equilibrium of an
ionized diatomic molecule. One only needs to take into account that
when a charged molecule dissociates, the atomic element which remains ionized
is the one with the lowest ionization potential:
\begin{equation}
\label{eq_dissociative_equilibrium_ionized}
AB^+ \Leftrightarrow A^+ + B,
\end{equation}
with $D_0(A)<D_0(B)$. Therefore, using Eq. (\ref{eq_equilibrium_constant}) and Eq. (\ref{eq_ionization_constant})
we can calculate the partial pressure of the ionized molecule:
\begin{equation}
K_p(T) = \frac{P_{A^+} P_B}{P_{AB^+}} = \frac{K_{A^+}(T)}{P_{e^-}} \frac{P_{A} P_B}{P_{AB^+}},
\end{equation}

We can write an equation like Eq. (\ref{eq_conservation_mass_2}) for each of the $N_s$ atomic species included in the calculation. In our
case we have selected the 21 species which are shown in Table \ref{tab_hidden_neurons}. Therefore, once
the molecular and ionization equilibrium constants are known, we have a set of $N_s$ non-linear algebraic equations which depend on
the local temperature and density. The ICE approximation is intrinsically local since it depends only on the local values of the
temperature and density. Therefore, to calculate the molecular number densities in a model atmosphere, we have to solve
the algebraic non-linear system of equations at each point in the
atmosphere. This system is solved using a standard Newton-Raphson
iterative method \citep{numerical_recipes86}.
Once the partial pressures for all the atomic species are known,
we can calculate the ensuing atomic number densities by using Eq. (\ref{eq_ideal_gas}). The molecular number densities are
calculated by solving for their partial pressures in Eq. (\ref{eq_equilibrium_constant}), using the equilibrium constant
appropriate for each molecule.

\subsection{Equilibrium constants}
As we discussed above, the problem of obtaining the atomic and molecular number densities is completely defined once the temperature
and the total density (or pressure) is known. However, we also need to know the value of the equilibrium constant for each
temperature. It is known from considerations of statistical mechanics that the value of the equilibrium constant can be obtained with
the aid of the partition function of the molecule and of the individual atomic components, their respective masses and the
dissociation energy $D_0$ \citep[see, e.g.,][ for detailed information on how the equilibrium constants can be calculated]{tejero_ordonez91}.
The expression is given by \citep[][ and references therein]{tejero_ordonez91}:
\begin{align}
\label{eq_equil_cte}
K_p(T) & = (k T)^{NA-1} \left( \frac{2 \pi k T}{h^2} \right) ^{3(NA-1)/2} \left( \frac{m_A^a m_B^b m_C^c}{m_{ABC}} \right) ^{3/2} \nonumber \\
&\times \left( \frac{\phi_A^a \phi_B^b \phi_C^c}{\phi_{ABC}} \right) e^{-D_0/kT},
\end{align}
where $m_i$ is the mass of atomic or molecular species $i$, $NA=a+b+c$ is the number of atoms present in the molecule, $\phi_i$ is
the partition function of species $i$. $h$ and $k$ stand for the Planck and Boltzmann constants, respectively. The partition functions
can be obtained by performing the summation
\begin{equation}
\label{eq_partition_function}
\phi = \sum_j g_j e^{-E_j/kT},
\end{equation}
where $g_j$ is the degeneracy of level $j$ and $E_j$ its energy. Ideally, the sum has to be extended over all the energy levels of the
species. However, due to the difficulty of accounting for all the energy levels, the most complete available
set of energy levels must be used. For molecules, the partition function has to include the contribution from the electronic, vibrational
and rotational levels. It is not practical to compute this summation for every
ICE calculations. As a workaround, some authors have tabulated polynomial fits
to the partition functions
\citep{russell34,tsuji71,sauvaltatum84,tejero_ordonez91}. In this work, we
make use of the results obtained by \citet{tejero_ordonez91}.
It represents one of the most complete compilations of molecular equilibrium constants for more than 240 diatomic and polyatomic molecules. These
equilibrium constants were obtained by calculating the partition function with the most up-to-date molecular constants at the
time of the
work and their dependence with the temperature are given, as in previous works, as polynomial fits. One of the advantages of selecting this
database is its self-consistency.

\section{Neural network}
\label{sec_ANN}
\subsection{Description of the network}
The ICE problem consists on obtaining the partial pressures of the atomic species (or the atomic species number densities) which are
consistent with Eq.~(\ref{eq_conservation_mass_2}) once the local temperature, local hydrogen density and local electron density are given.
One could also take a different point of view and see the ICE problem as a
mapping between
the three-dimensional space
given by $(T,n_H,n_e)$ onto several one-dimensional spaces, one for each atomic species included in the problem. The functional form
of the functions $f : \mathbf{R}^3 \to \mathbf{R}$ is not known.

\begin{figure}
\resizebox{\hsize}{!}{\includegraphics{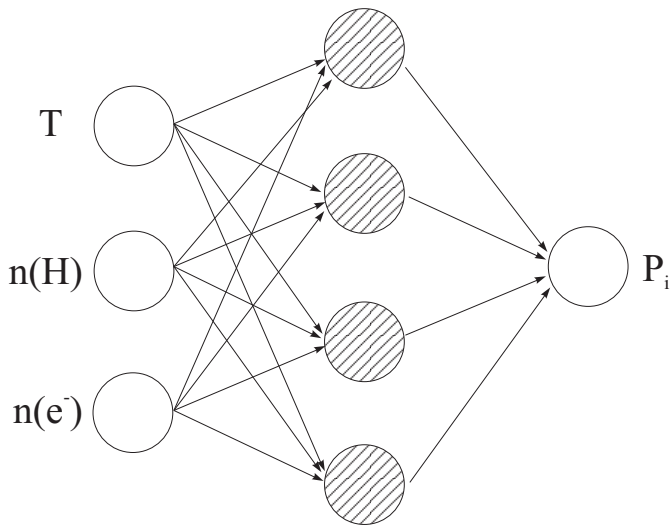}}
\caption{Schematic structure of the neural network model used for approximating the ICE results. The input layer and the output layer
are linear, while the hidden layer is assumed to be non-linear, with an hyperbolic-tangent activation function. The input layer consists
of three neurons for the temperature, hydrogen number density and electronic number density. The output neuron gives the partial pressure
of element $i$ for the physical conditions in the input layer.}
\label{fig_ANN}
\end{figure}

The Artificial Neural Network (ANN) with one hidden layer is
a universal approximant to any non-linear continuous function \citep[e.g.,][]{jones90,blum91}.
The schematic structure of such a network is shown in Fig.
\ref{fig_ANN}. We have constructed an ANN with an input layer of three
neurons where the temperature, hydrogen density and
electron density are introduced. The neurons of the input layer have a linear activation function. The hidden layer consists on $N_h$ neurons
with a non-linear activation function $\sigma(x)$. Each hidden neuron is
connected to all the neurons of the input layer by a certain weight
(indicated by arrows in the figure). The value obtained at each hidden neuron is a linear combination of the values
at the neurons of the input layer multiplied by the weights. These values are then applied the non-linear function $\sigma(x)$,
multiplied by another set of weights and summed to give the final output of the neural network. The output of the network can be written
as:
\begin{equation}
N(T,n(\mathrm{H}),n(\mathrm{e}),\vec{m}) = \sum_j^{N_h} v_j \sigma \left[ w_j^\mathrm{T} T + w_j^\mathrm{H} n(\mathrm{H}) +
w_j^\mathrm{e} n(\mathrm{e}) + u_j\right],
\label{eq_neural_network}
\end{equation}
where the activation function is usually given by the sigmoid or the hyperbolic-tangent function. In this case, we have selected
$\sigma(x)=\tanh (x)$. In the previous expression, $\vec{m}$ formally
represents the whole set of $5 \times N_h$ weights which define the
neural network.

The use of a neural network approach for the solution of the ICE problem is very appealing
for several reasons.
Firstly, we obtain an analytic function
which is infinitely differentiable and which could be easily used in any subsequent calculation. This way, we can investigate with great
detail the dependence of the atomic and/or molecular abundances with temperature, hydrogen density and electron density. Secondly, the
powerful approximation capabilities of ANNs make them very suitable for
multi-dimensional interpolation problems. Even more striking is the ability
of ANNs to extrapolate data outside the range of parameters used in the
learning process. A direct consequence of these
properties is that the number of points that need to be used for the learning
process can be very small compared to
the case of standard interpolation techniques. Furthermore, the number of parameters needed for performing the interpolation
is also small. Essentially, one transforms the problem from storing in memory the whole set of $N$ points which relates $T$, $n(H)$ and $n(e)$ with
the partial pressures $P_i$ for every species to storing only the
$5 \times N_h$ $\vec{m}$ weights present in Eq. (\ref{eq_neural_network}) for each
species. This reduction is proportional to the ratio $N/N_h$, which can be very large for large, as we will see
below. Finally, it is also important to note that the neural network can be straightforwardly implemented in parallel architectures.

\subsection{Learning process}
In order to obtain a neural network which is able to approximate the abundance of each species included in the chemical equilibrium
calculation, we randomly select $N_l$ sets of temperature, hydrogen density and electronic density. For each one of these combinations,
we solve the ICE equations and calculate the abundances of all the atomic species. We have verified that taking $N_l=1000$ gives a good
coverage of the parameter space, considering that they are randomly selected. The temperature is varied in the range from
3500 K and 14000 K, which is representative of the physical conditions in typical stellar atmospheres. Concerning the
hydrogen and electronic abundances, we decided to use a logarithmic scaling.
The hydrogen density
is varied from 10$^{12}$ cm$^{-3}$ to 10$^{18}$ cm$^{-3}$ while the electronic density is varied from 10$^8$ cm$^{-3}$ to 10$^{17}$ cm$^{-3}$.
Again, these values are a reasonable representation of realistic conditions.

Once the learning set is selected, the training of the network for each species $i$ reduces to a minimization of the following error function:
\begin{equation}
E_i = \sum_{l=1}^{N_l} \left[ P_i^l - N(T_l,n_l(\mathrm{H}),n_l(\mathrm{e}),\vec{m}) \right]^2,
\label{eq_error_training}
\end{equation}
so that the optimum values of the parameters $\vec{m}$ are obtained. $P_i^l$ is the partial pressure (or number density) of species $i$
obtained for the combination of physical conditions $l$. Since almost every minimization algorithm makes use of the derivatives
of the error function with respect to the parameters, it is of interest to notice that they can be obtained analytically:
\begin{equation}
\frac{\partial E}{\partial m_k} = -2 \sum_{l=1}^{N_l} \left[ P_i - N(T_l,n_l(\mathrm{H}),n_l(\mathrm{e}),\vec{m}) \right]
\frac{\partial N}{\partial m_k},
\label{eq_derivatives}
\end{equation}
where the partial derivatives of the neural network with respect to the parameters are obtained in a straightforward manner by using
Eq. (\ref{eq_neural_network}). In order to obtain the set of parameters $\vec{m}$ which minimize the error function, we have used a
quasi-Newton BFGS method \citep{fletcher87} implemented in the Merlin package \citep{merlin98} which presents a very good convergence rate.

\begin{figure*}
\includegraphics[width=8cm]{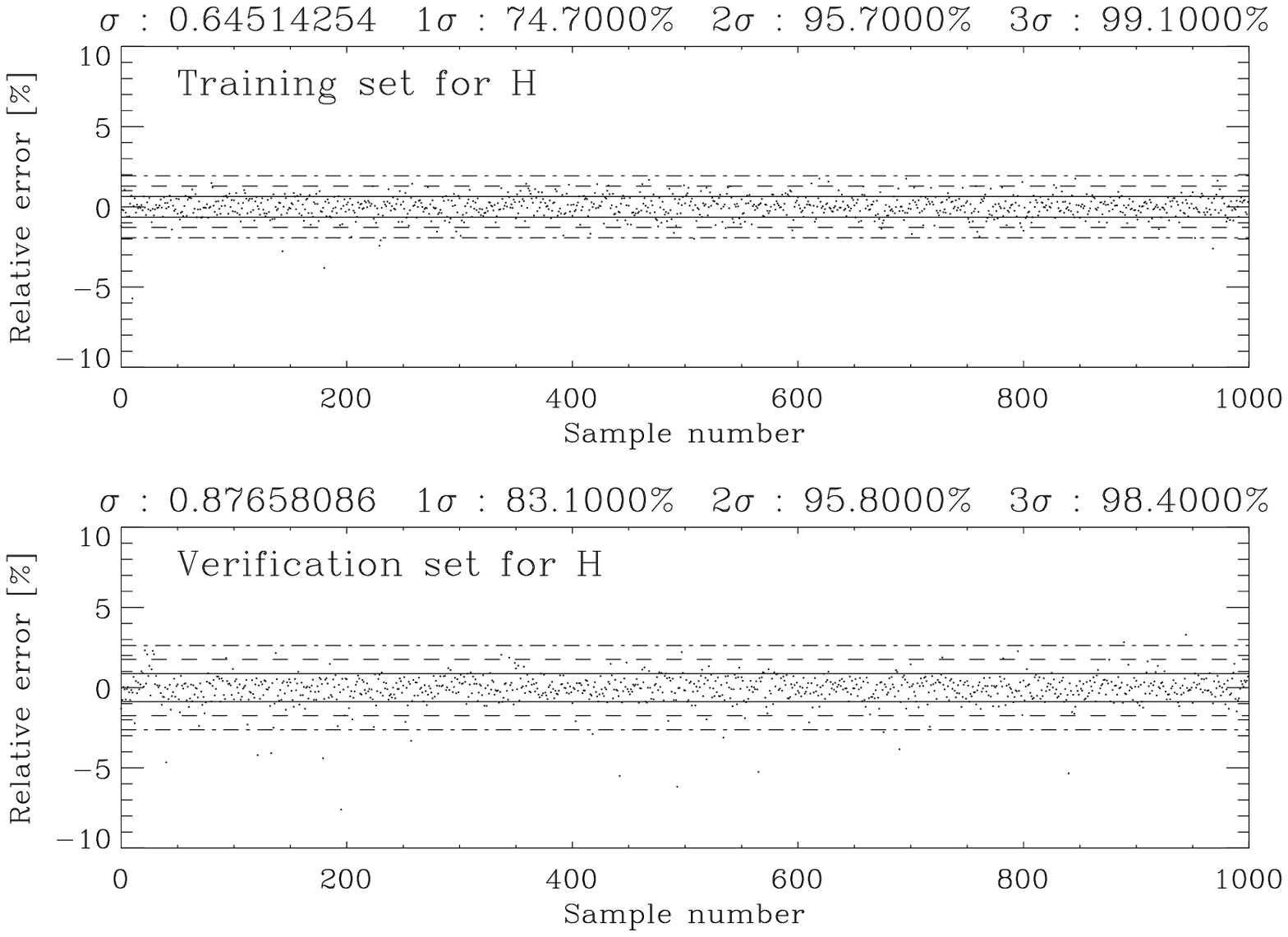}\hspace{1cm}%
\includegraphics[width=8cm]{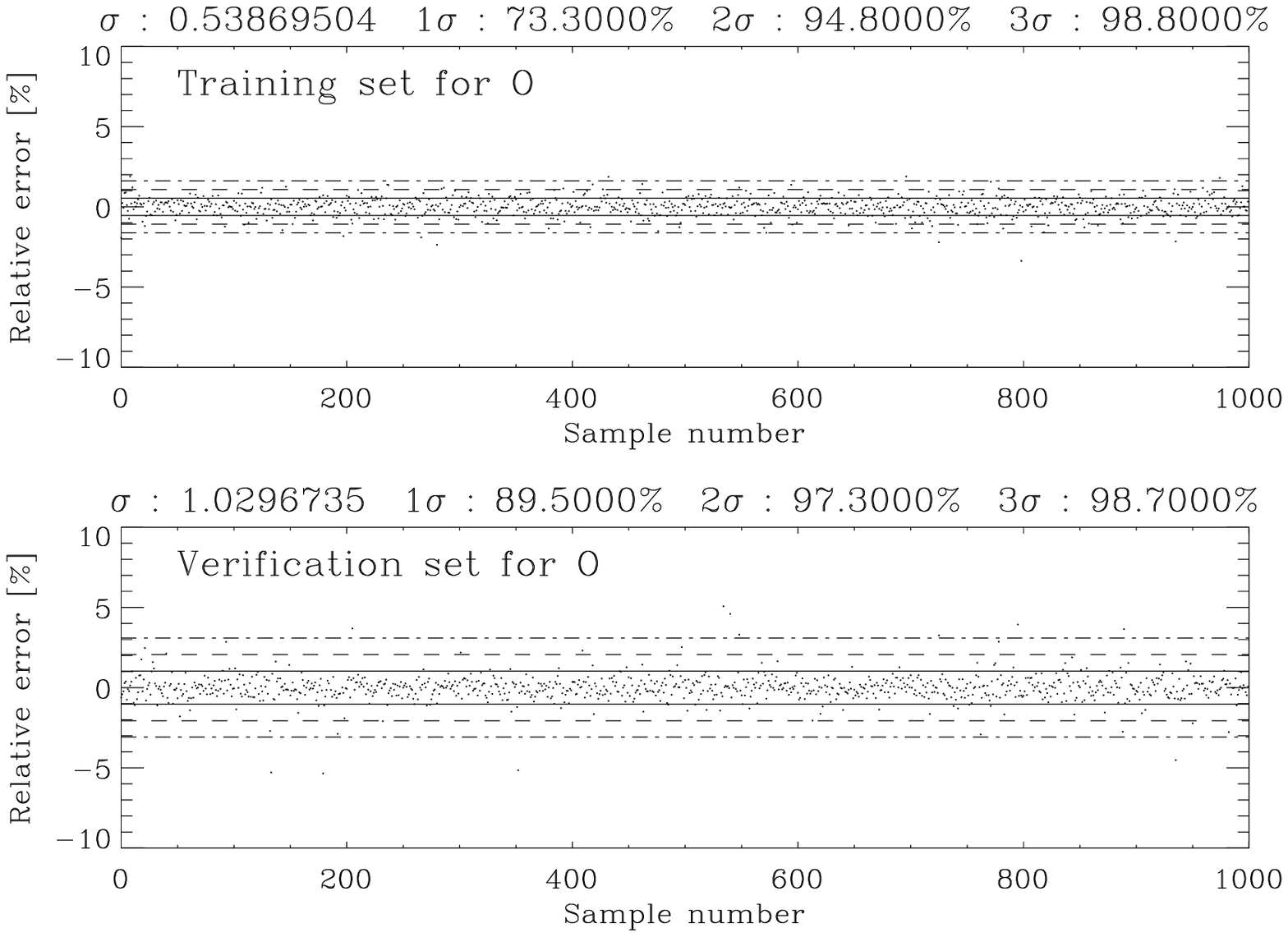}
\includegraphics[width=8cm]{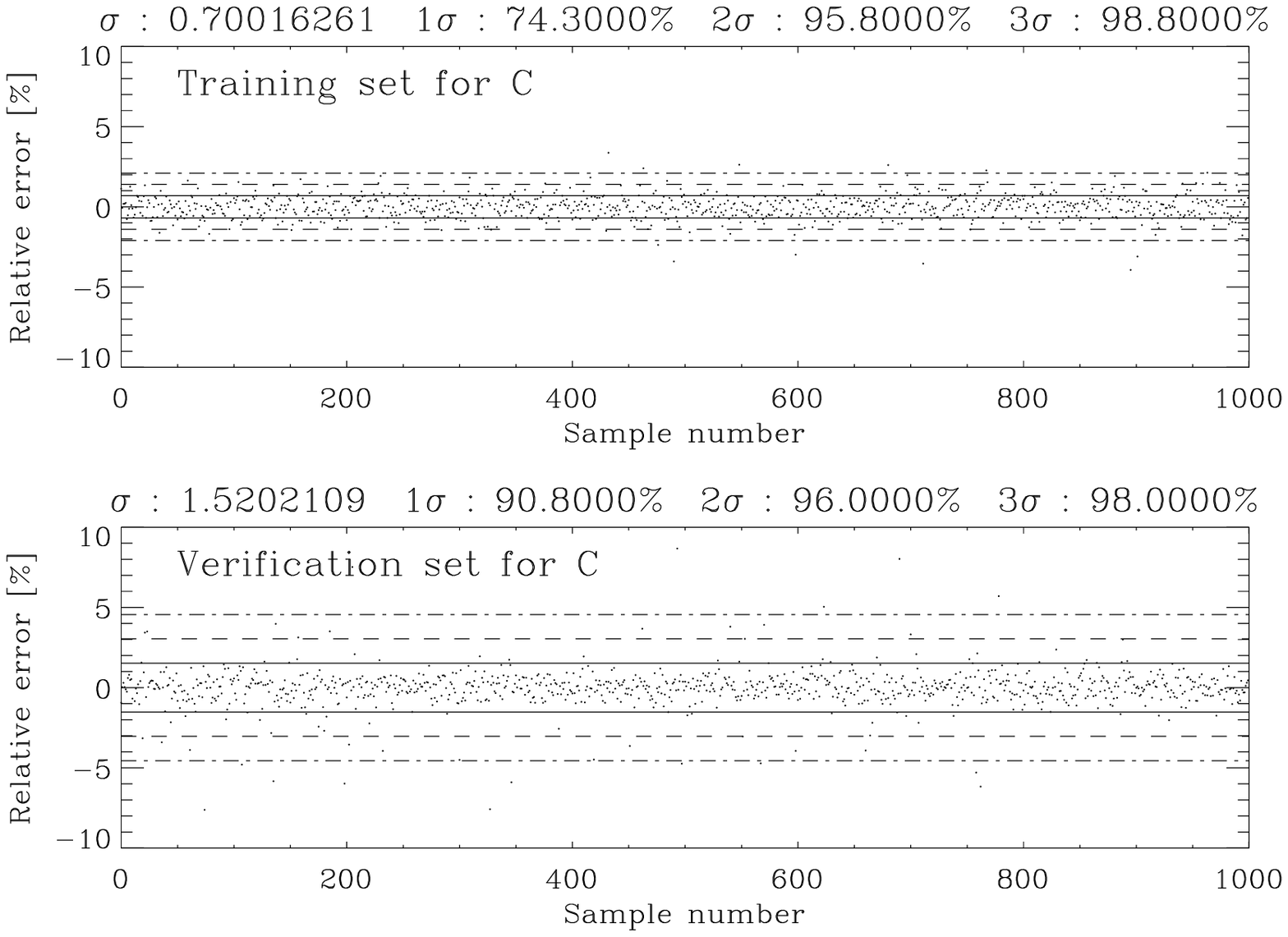}\hspace{1cm}%
\includegraphics[width=8cm]{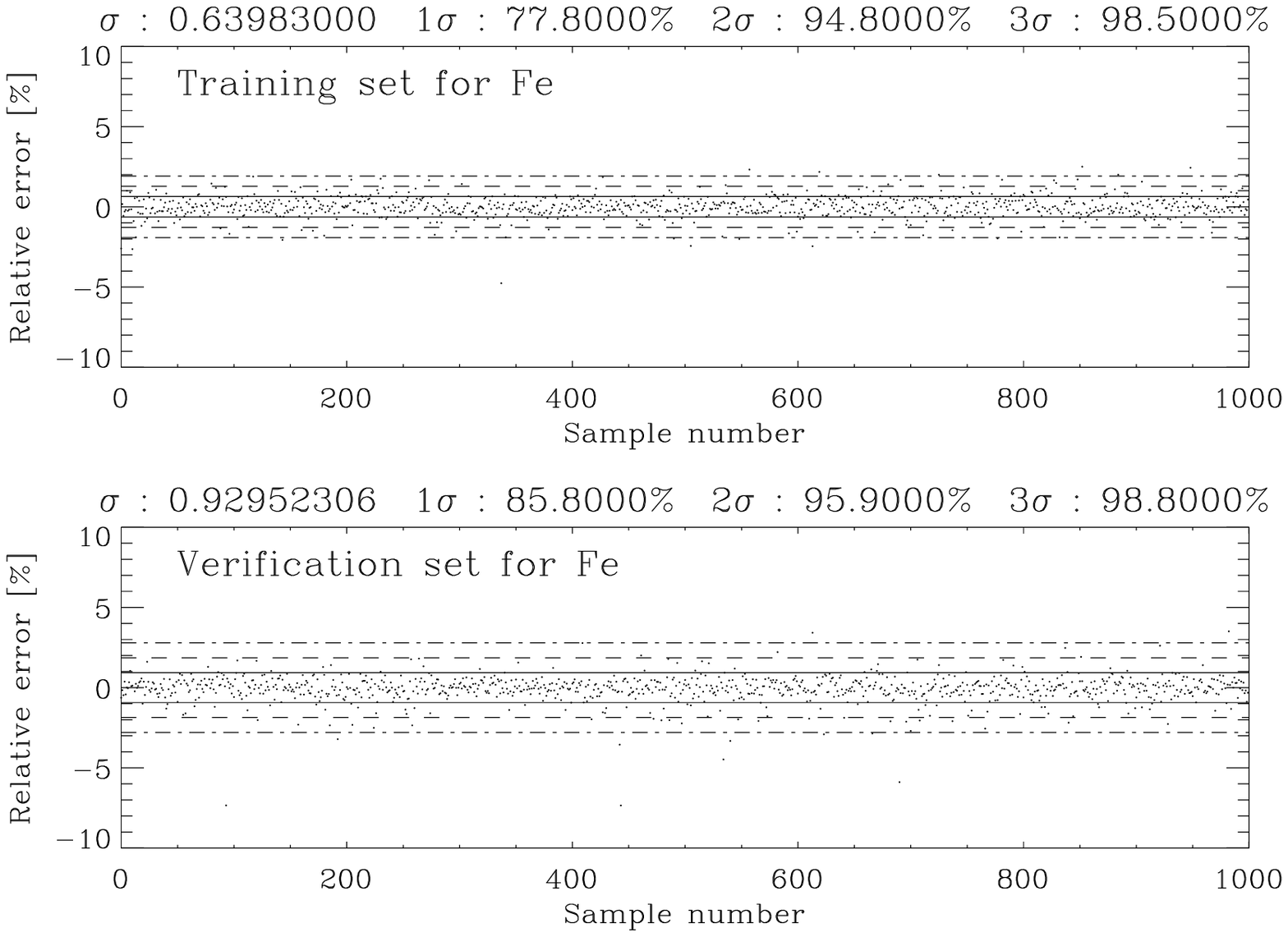}
\caption{Relative error obtained after the training of the neural network for hydrogen, oxygen, carbon and iron. The distribution of
relative errors are shown for the learning set and for a verification set, which aids at detecting overtraining. We indicate the value
of the standard deviation $\sigma$ of the distribution and the number of points below 1$\sigma$, 2$\sigma$ and 3$\sigma$. We also indicate
these values in the plots as horizontal lines.}
\label{fig_training1}
\end{figure*}

It is interesting to note that the usual approach of training the neural network with a learning set obtained from the original problem
can be circumvented here. We do not need to rely on the precision given by the
Newton-Raphson solution and the neural network can be made as precise as desired. This is a consequence of the fact that the equations
describing the ICE approximations are known. One can see that the equations of ICE can be formally written as $f(T,n(H),n(e),\{P_i\})=0$,
where $\{P_i\}$ is the set of $N_s$ partial pressures (or atomic number densities) which satisfy the equations. Therefore, we can build
as many neural networks as species included in the ICE problem and train the network by minimizing the following error function:
\begin{equation}
E = \sum_{l=1}^{N_l} \left[ f(T_l,n_l(\mathrm{H}),n_l(\mathrm{e}),\{ N_i(T_l,n_l(\mathrm{H}),n_l(\mathrm{e}), \vec{m}) \}) \right]^2.
\label{eq_error_training2}
\end{equation}
The solution to the problem is transformed into finding the set of parameters $\vec{m}$ for all the networks which minimizes the previous
error function. The summation is extended over the $N_l$ elements of the learning set. Several issues are noteworthy in this case.
On the one hand, the derivatives of the error function with respect to the parameters $\vec{m}$ are not as straightforward as those found
in Eq. (\ref{eq_derivatives}) since the function $f(T,n(H),n(e),\{P_i\})=0$ is non-linear. On the other hand, the minimization process
has to be carried out simultaneously for the $N_s$ neural networks, so that the problem of finding the minimum might be more difficult
to solve. The dimension of the hypersurface in which we have to find the
minimum is $5 \times N_h \times N_s$, instead of finding $N_s$ minima
in hypersurfaces of dimension $5 \times N_h$.

A critical parameter in the training ability of a neural network is the number of hidden neurons $N_h$. It is important to build
networks with sufficient number of hidden networks to accurately approximate the non-linear function we are interested
in. However, since the number of learning points is not infinite, the number of hidden neurons one can use is practically limited by the
overtraining phenomenon (in principle, the number of training points should be much larger than the number of hidden neurons). If $N_h$
is very large, the network is capable of correctly approximating all the points in the learning
set but its interpolation capacity is lost, producing large oscillations between the points of the learning set. This is a direct
consequence of the increase in the degrees of freedom of the network when the number of hidden neurons is increased. One way to avoid
this overtraining is to use two different data sets: one for learning purposes and the other for testing. When the training process
takes places, the error given by expression (\ref{eq_error_training}) is evaluated for the training set and for the test set. When
the network is training correctly, both errors are reduced. When the network starts to be overtrained, the error of the test set
starts to increase. We have followed this scheme in order to stop the
training process before overtraining.
In our case, we have used two different values for the number of hidden neurons. We have verified that $N_h=20$ gives sufficiently
accurate results for 13 of the 21 atomic species included in the ICE calculations, while we were forced to increase this number to
$N_h=30$ for the rest of species. Table \ref{tab_hidden_neurons} lists the
number of hidden neurons for each species. We
did not investigate the dependence of the minimum error on the number of hidden neurons so that the increase to $N_h=30$ was
somewhat arbitrary.

\begin{table}
\caption{Number of hidden neurons used for each ANN.}
\label{tab_hidden_neurons}
\centering
\begin{tabular}{c c}
\hline\hline
Species & Number of hidden neurons \\
\hline
He, Na, Mg, Al, P, K, Ca & 20 \\
Ti, Cr, Mn, Fe, Ni, Cu & \\
H, C, N, O, F, Si, S, Cl & 30 \\
\hline
\end{tabular}
\end{table}

\begin{figure*}
\includegraphics[width=8cm]{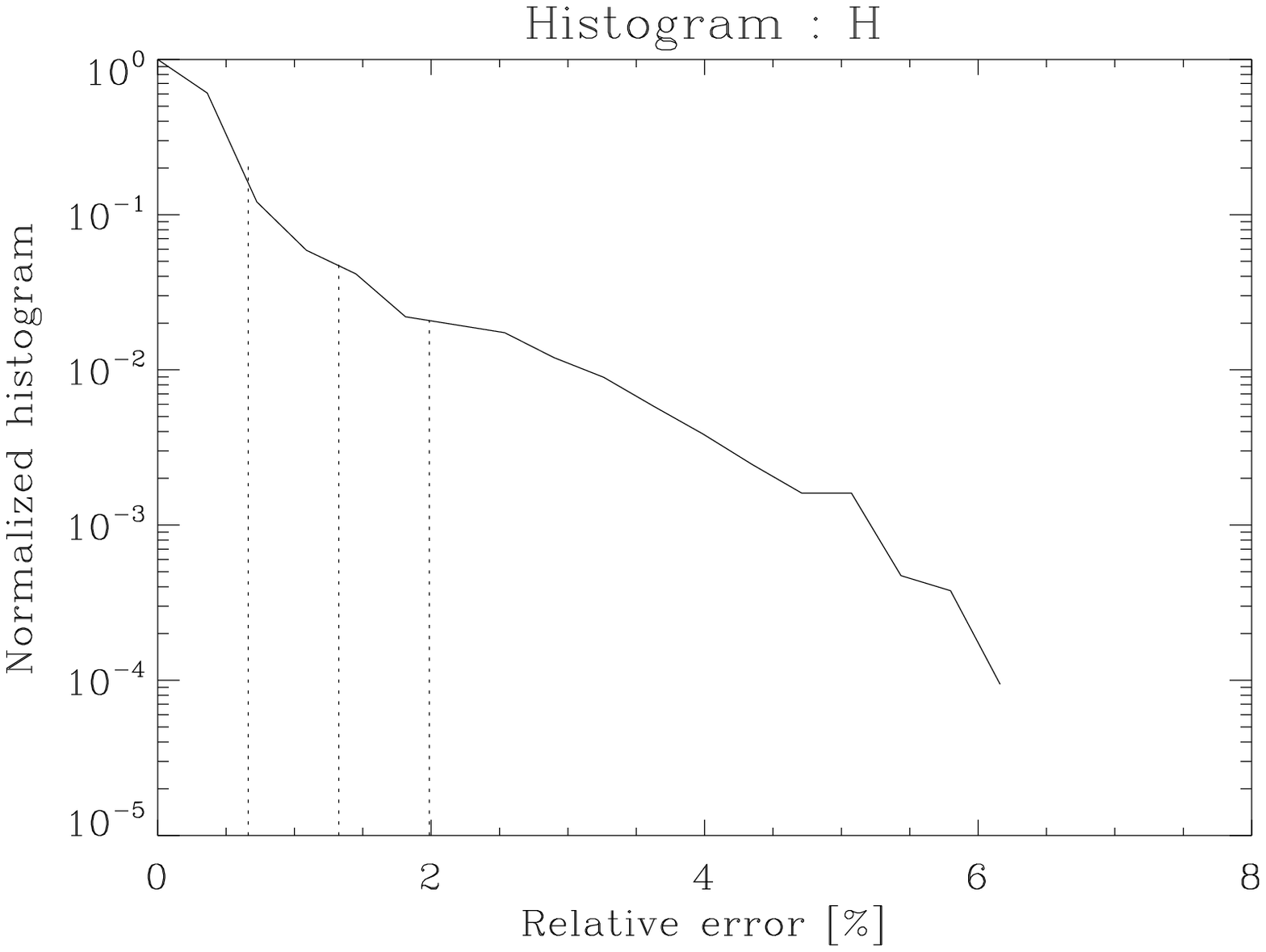}\hspace{1cm}%
\includegraphics[width=8cm]{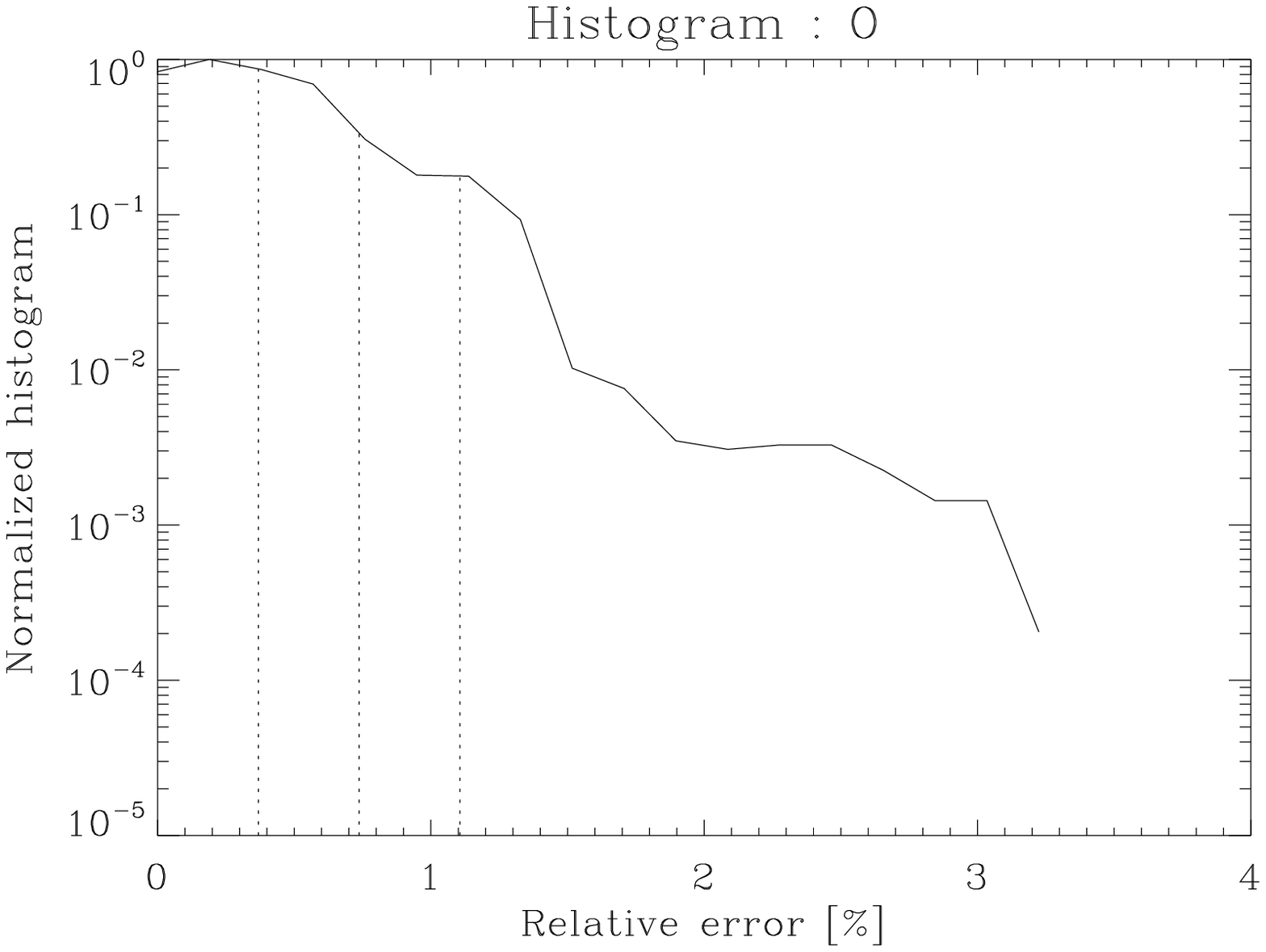}
\includegraphics[width=8cm]{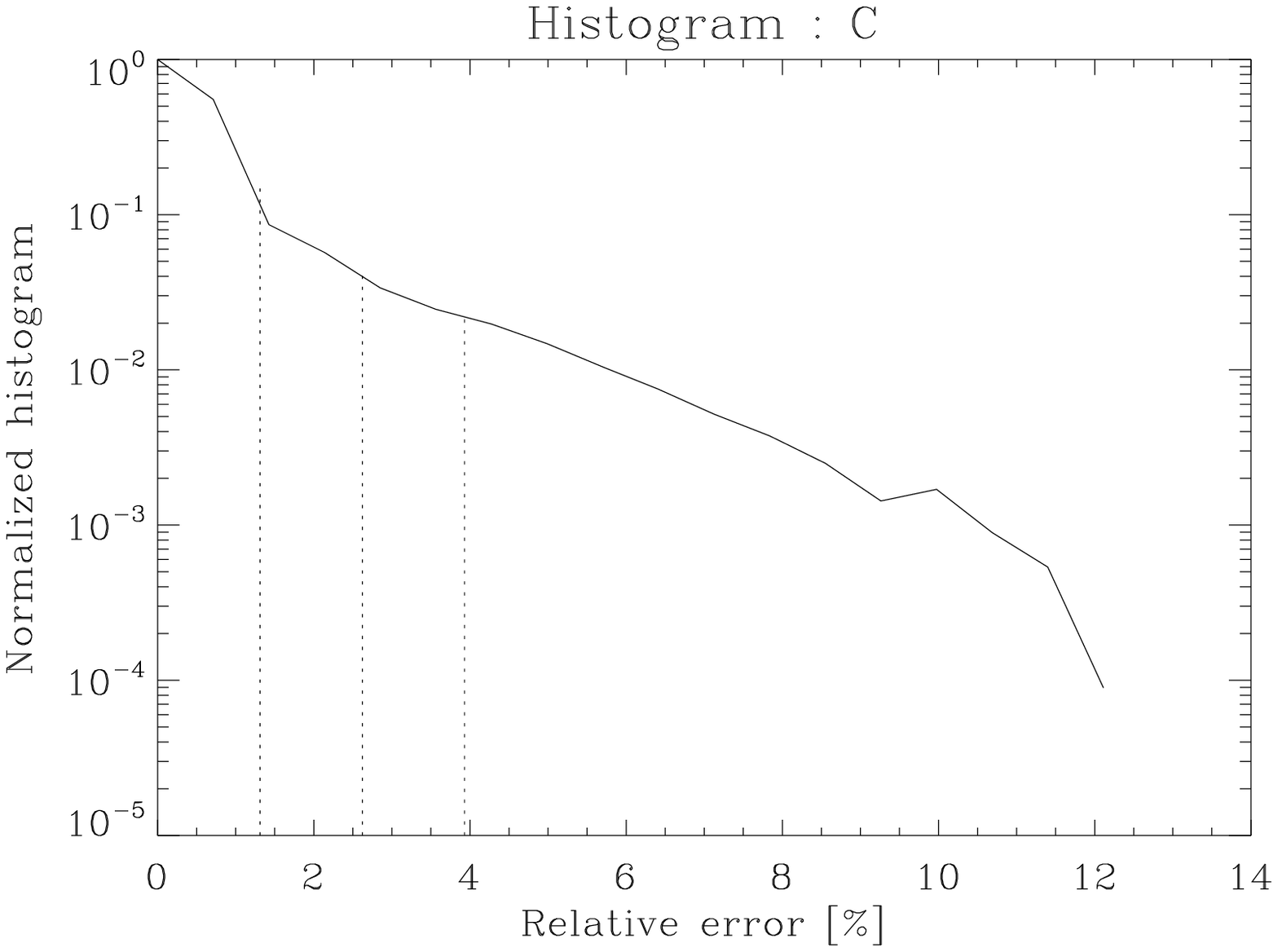}\hspace{1cm}%
\includegraphics[width=8cm]{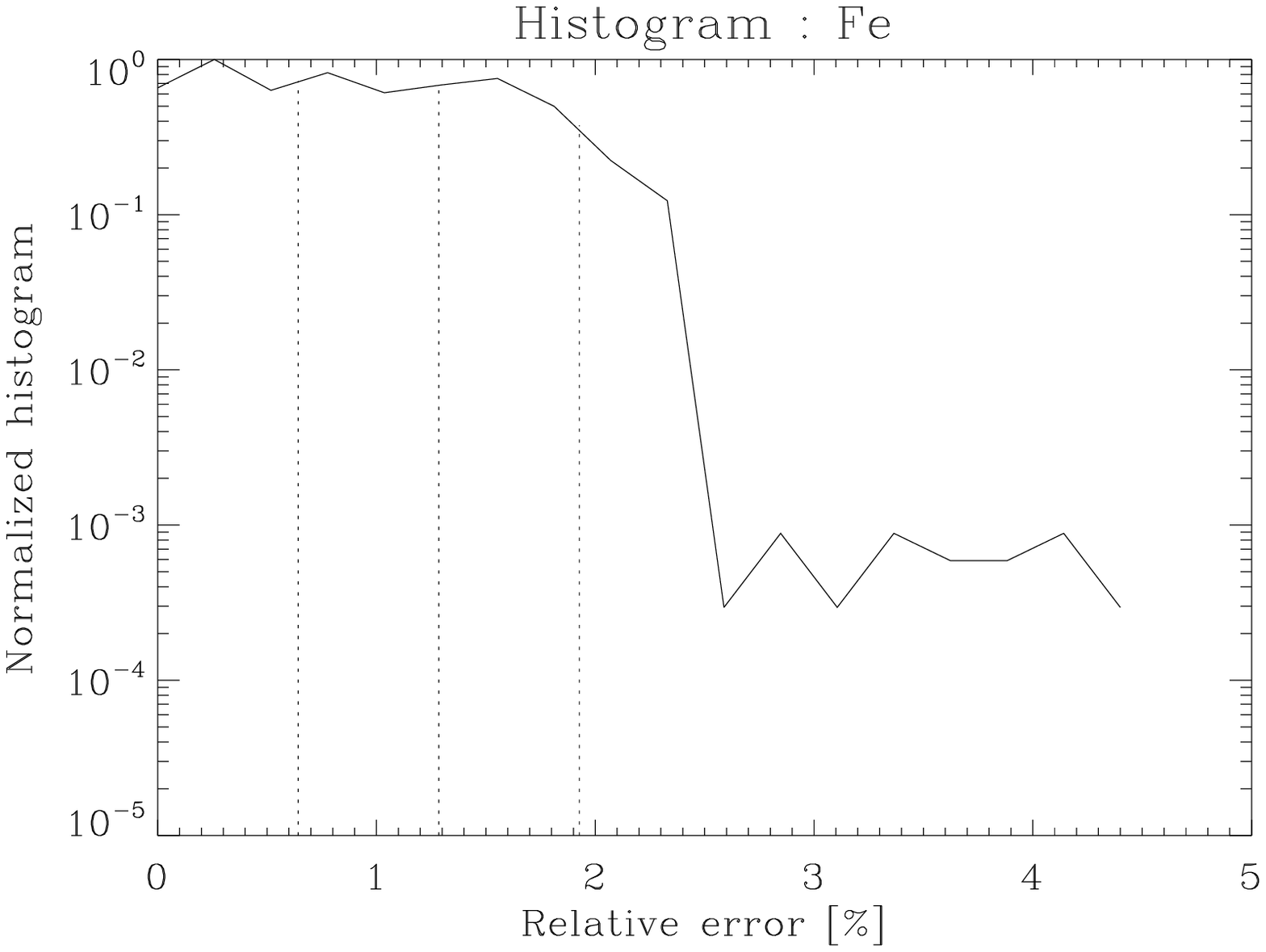}
\caption{Histograms of relative errors for the 2.55$\times$10$^5$ points of the three-dimensional convection simulation of
\citet{asplund00}. We indicate the position of 1$\sigma$, 2$\sigma$ and 3$\sigma$ of the distribution. Note that they are similar to
those obtained for the verification set during the training of the networks.}
\label{fig_comparison}
\end{figure*}

The training of the neural networks was stopped when the standard deviation of the relative error between the output of the neural network
and the values obtained
from the solution of the ICE equations for both the learning and the test sets was 1\% or below. Obviously, the relative error
was always larger for the test set than for the learning set, but the training always reached the limit of 1\% and was stopped
before any indication of overtraining was observed. In Fig. \ref{fig_training1} we present the relative error obtained for the 1000
training points for the abundances of H, C, O and Fe. The behavior of the relative error for the rest of species follows the same
behavior, except for He, which presents a much smaller dispersion. Since He cannot form molecules, it is always in its atomic form and
the non-linear mapping that the neural network has to learn is strongly simplified.

We have indicated in each plot the value of the standard deviation and the
percentage of points which are contained within 1$\sigma$,
2$\sigma$ and 3$\sigma$. This is an indication of the precision obtained with
the neural network. It is common in all the networks
that more than $\sim 95$\% of the points are inside 3$\sigma$, which translates into a relative error of $\sim 3$\% (since $\sigma$ is
always of the order or below 1\%). However, we note that it is common to all the networks that more than 75\% of the points are
below 1$\sigma$. Along with the plots of the results obtained for the learning set, we have included the results for the verification
set. It is interesting to note that, although the standard deviation of the distribution of relative errors is larger than that obtained
for the learning set, typically more than $\sim 80$\% of the points are below this limit. In the case of the network for the carbon
abundance, we have a relatively high value for $\sigma$, but more than 90\% of the points are below this limit.

\section{Comparison of results}
\label{sec_comparison}
Once the neural networks were trained to the desired precision, as shown in Fig. \ref{fig_training1}, we have applied
them to realistic situations in order to compare their performance with the
complete ICE calculations. To this end, we make use of a
three-dimensional snapshot of the convection simulation of the solar atmosphere of \citet{asplund00}.
We consider this simulation to be
an adequate representation of the physical conditions in the solar atmosphere since several investigations are reporting good agreements
between the synthetic line profiles and the observed ones \citep[e.g.,][]{shchukina_trujillo01,asplund_oh_03,asplund_li_03}. The snapshot box
size is 50x50x102, so that the number of points is 2.55$\times$10$^5$. We have solved the ICE problem in each point
and confronted the results with those obtained from
the neural networks.

\begin{figure*}
\includegraphics[width=8cm]{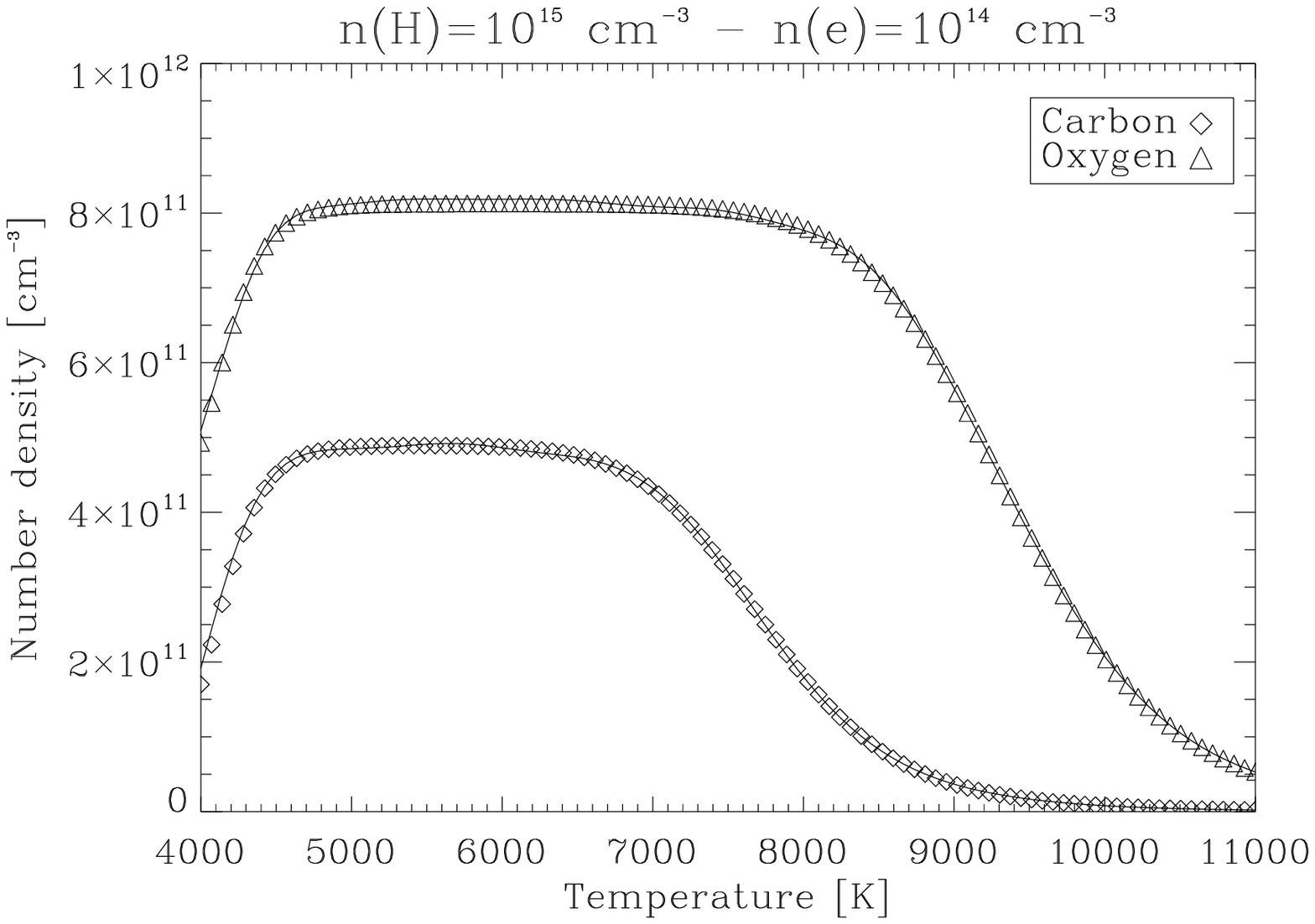}\hspace{1cm}%
\includegraphics[width=8cm]{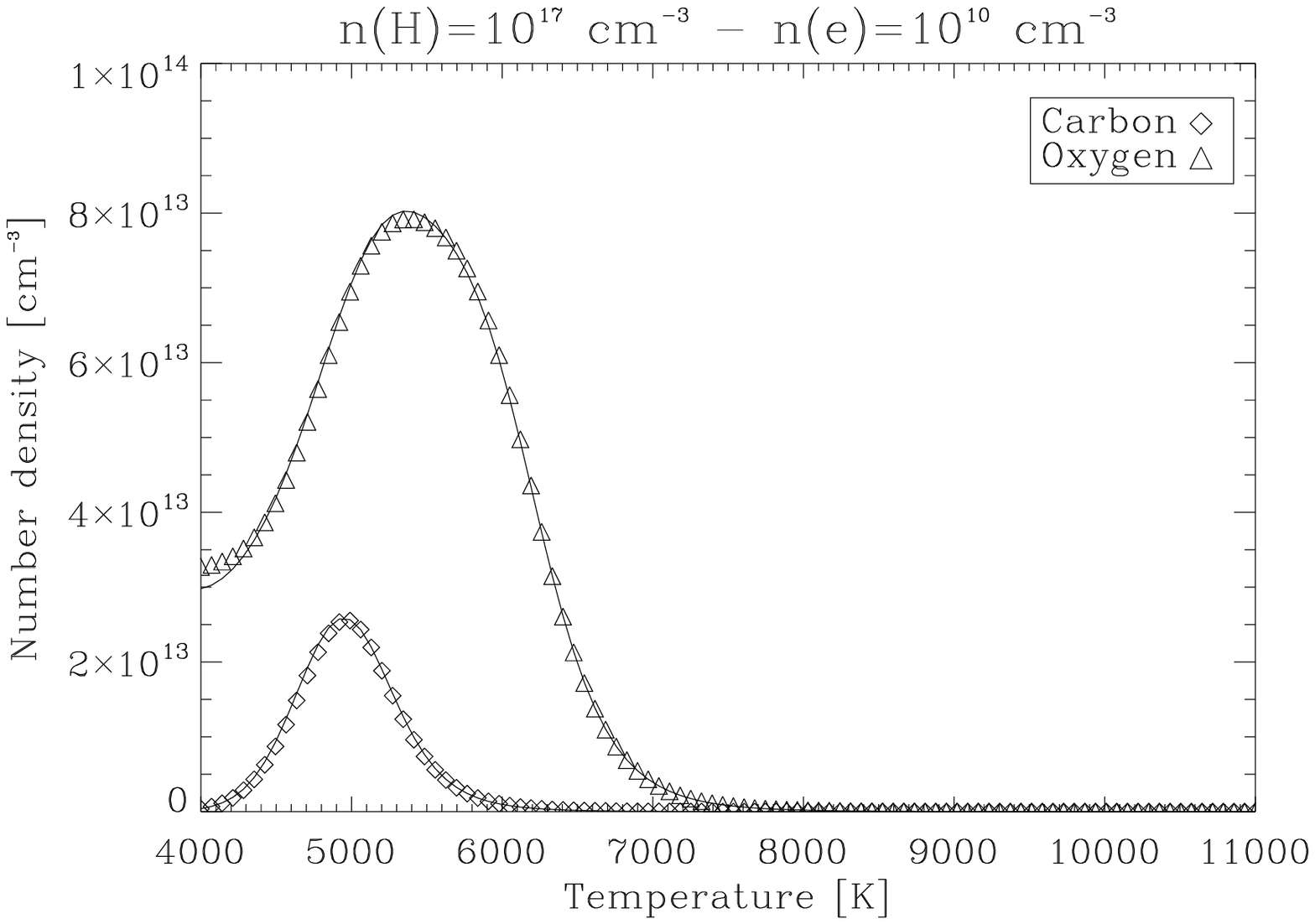}
\caption{Number density of carbon and oxygen obtained with the simplified model and with the neural network for two different combinations
of hydrogen and electron densities and for different values of the temperature. Note that this kind of investigation is greatly
simplified due to the analytical character of the neural network.}
\label{fig_simplified}
\end{figure*}

The normalized error histograms are shown in Fig. \ref{fig_comparison} for H, C, O and Fe, with vertical dashed lines indicating the position
of 1$\sigma$, 2$\sigma$ and 3$\sigma$ of the distribution. Note that the histograms are decreasing functions of the relative error, so that
the number of points with large relative error is much smaller than the number of points with small relative errors. It is verified that
the value of the standard deviation of the distribution of relative errors is quite similar to those obtained for the verification
set. This reinforces our confidence that the learning set and the verification set
provide sufficient coverage of the three-dimensional parameter
space $(T,n(H),n(e))$. The shapes of the histograms turn out to be fairly linear in this log-linear scale, so that we can consider, as a first
approximation, that the number of points $N_p$ with a certain relative error $r$ is given by $N_p \propto r^{-\alpha}$. For example, we find
that $\alpha \simeq 5$ for hydrogen, $\alpha \simeq 3.7$ for carbon and $\alpha \simeq 7$ for oxygen. There is an exception for iron, for
which the distribution of points with relative errors below $\sim 2.3$\% seems to be constant, while the number of points with errors
larger than $\sim 2.3$\% is three order of magnitude smaller. The behavior of the histograms for the rest of species is very similar to
those shown in Fig. \ref{fig_comparison} for the selected ones, maintaining the properties obtained for the verification sets.

\begin{figure*}
\includegraphics[width=8cm]{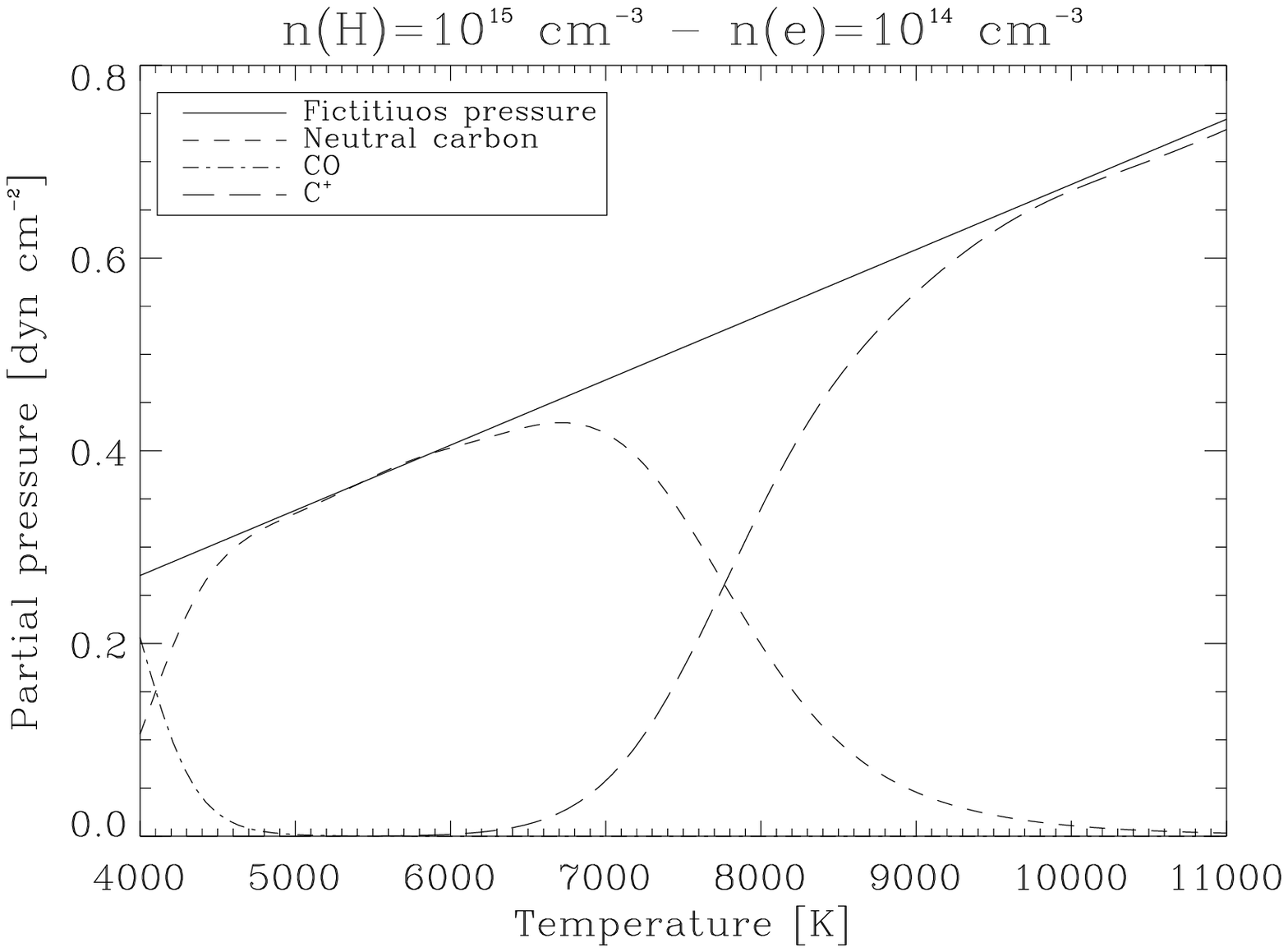}\hspace{1cm}%
\includegraphics[width=8cm]{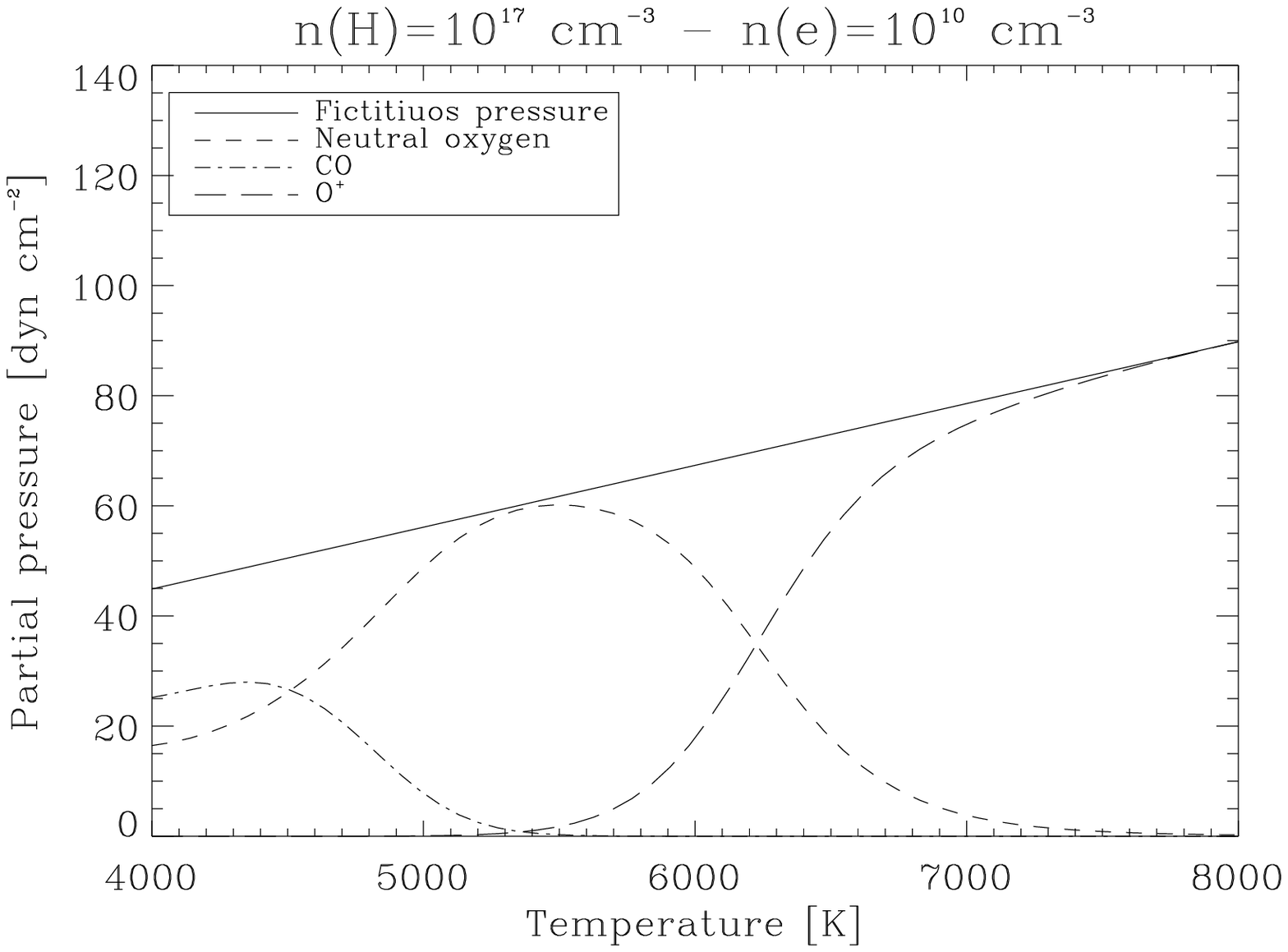}
\caption{Contribution to the total pressure of carbon and oxygen in two different cases. The curves have been obtained using the
neural networks. The addition of all the contribution closely matches the fictitious pressure of carbon and oxygen. This is the reason
why the very simplified model for C and O works very well.}
\label{fig_simplified2}
\end{figure*}

\section{$T$, $n(H)$ and $N(e)$ dependence}
\label{sec_dependence}
Since the neural network approach leads to analytic formulae for the solution of ICE problems, we now take advantage of them
and compare the results for abundant species with simplified models. Our aim is to investigate the variation with temperature, hydrogen
density and electronic density of the number density of carbon and oxygen. We have selected these two species because they are
the constituents of the highly abundant CO diatomic molecule. This way, we can safely obtain an approximation to the carbon and oxygen
abundances by taking only into account the formation of CO in the ICE equations and neglecting the rest of molecular species. Note that
this simplification is only valid for species like C and O and not for species which form less abundant molecules and/or form several
high abundance molecules. Specializing Eq. (\ref{eq_conservation_mass_2}) to oxygen and carbon, we obtain:
\begin{align}
P(C) &= P_C + \frac{P_C P_O}{K_{CO}(T)} + K_{C^+} \frac{P_C}{P_{e^-}} + K_{C^-} P_C P_{e^-} \nonumber \\
P(O) &= P_O + \frac{P_C P_O}{K_{CO}(T)} + K_{O^+} \frac{P_O}{P_{e^-}} + K_{O^-} P_O P_{e^-}.
\label{eq_ICE_simplified}
\end{align}
This system of equations can be solved analytically for $P_C$ and $P_O$ and transformed into number densities by using the ideal gas
equation Eq. (\ref{eq_ideal_gas}). Their dependence with temperature for two different combinations of hydrogen and electronic densities
are shown in Fig. \ref{fig_simplified}. The output of the neural network is also shown in the plots with solid lines. Note the accurate
results obtained with this simplified model in this high abundance case. This behavior can be better understood if we plot separately the
contribution to the total pressure of each element of each term in Eq. (\ref{eq_ICE_simplified}). The results are shown in Fig.
\ref{fig_simplified2} for the two sets of physical conditions. The fictitious pressure of carbon and oxygen (essentially the abundance of
carbon and/or oxygen times the hydrogen density) is shown by solid line, the partial pressure of the neutral
element by dashed line, the
partial pressure of the carbon monoxide molecule by dash-dotted line and the
partial pressure of the ionized element by long dashes. All the quantities
shown in the plots have been obtained using the neural network approach. Although the neural network takes into account the formation of
other molecules which contain carbon and oxygen (i.e., it gives the solution to the complete ICE problem), we see that accounting only
for CO appears to be enough since the addition of all the contribution closely approximates the value of the fictitious pressure.

Note that the CO molecule is efficiently formed when the temperature is below $\sim$5000 K in both situations. The neutral species are
mainly situated at intermediate temperatures, typical from photospheric regions, while the ionized species tend to dominate for high
temperatures, above $6000-7000$ K.

\section{Conclusion}
\label{sec_conclusion}
We have successfully trained 21 neural networks which approximate the solution to the Instantaneous Chemical Equilibrium problem. We have
generated a learning set which is representative of the physical conditions in stellar atmospheres and have verified that the coverage
of the three-dimensional parameter space is sufficiently good with only 1000 points. We have trained the neural networks so that the
standard deviation of the relative error
were below or of the order
of 1\%. In order to avoid overtraining, we have employed an independent
verification set.
The standard
deviation of the relative error in this verification set is larger than that for the learning set but also of the order of 1\%. We apply
the neural networks obtained to the ICE problem in a three dimensional
convection simulation, representative of the physical conditions
in the solar atmosphere. The histograms of relative errors built with all the points in the simulation show that the neural networks
are capable of solving the ICE problem with relative errors similar to those obtained for the verification set. The advantage of
the neural network approach is their intrinsic analytical character, the
possibility of parallelization and the very fast evaluation. The development of a fast approach to ICE is of importance for, among
others, recent multi-dimensional simulations of stellar atmospheres which include very dense grids and iterative inversions of
observed spectra.

\begin{acknowledgements}
\thanks{This research has been partly funded by the Ministerio de Educaci\'on y Ciencia through project AYA2004-05792 and by
the European Solar Magnetism Network (contract HPRN-CT-2002-00313).}
\end{acknowledgements}


\bibliographystyle{aa}
\bibliography{c:/Thesis/biblio}

\begin{thebibliography}{22}
\expandafter\ifx\csname natexlab\endcsname\relax\def\natexlab#1{#1}\fi

\bibitem[{{Asensio Ramos} {et~al.}(2003){Asensio Ramos}, {Trujillo Bueno},
  {Carlsson}, \& {Cernicharo}}]{asensio03}
{Asensio Ramos}, A., {Trujillo Bueno}, J., {Carlsson}, M., \& {Cernicharo}, J.
  2003, \apj, 588, L61

\bibitem[{{Asplund} {et~al.}(2003){Asplund}, {Carlsson}, \&
  {Botnen}}]{asplund_li_03}
{Asplund}, M., {Carlsson}, M., \& {Botnen}, A.~V. 2003, \aap, 399, 31

\bibitem[{{Asplund} {et~al.}(2004){Asplund}, {Grevesse}, {Sauval}, {Allende
  Prieto}, \& {Kiselman}}]{asplund_oh_03}
{Asplund}, M., {Grevesse}, N., {Sauval}, A.~J., {Allende Prieto}, C., \&
  {Kiselman}, D. 2004, \aap, 417, 751

\bibitem[{{Asplund} {et~al.}(2000){Asplund}, {Ludwig}, {Nordlund}, \&
  {Stein}}]{asplund00}
{Asplund}, M., {Ludwig}, H.~G., {Nordlund}, A., \& {Stein}, R.~F. 2000, \aap,
  359, 669

\bibitem[{Bishop(1996)}]{B96}
Bishop, C.~M. 1996, Neural networks for pattern recognition (Oxford University
  Press)

\bibitem[{{Blum} \& {Li}(1991)}]{blum91}
{Blum}, E.~K. \& {Li}, L.~K. 1991, Neural Networks, 4, 511

\bibitem[{{Carroll} \& {Staude}(2001)}]{carroll01}
{Carroll}, T.~A. \& {Staude}, J. 2001, \aap, 378, 316

\bibitem[{{Fletcher}(1987)}]{fletcher87}
{Fletcher}, R. 1987, Practical Methods of Optimization, 2nd ed. (New York:
  Wiley)

\bibitem[{{Grevesse}(1984)}]{grevesse84}
{Grevesse}, N. 1984, \physscr, 8, 49

\bibitem[{{Jones}(1990)}]{jones90}
{Jones}, L.~K. 1990, in Proceedings of the IEEE, 78, 1585

\bibitem[{{McCabe} {et~al.}(1979){McCabe}, {Connon Smith}, \&
  {Clegg}}]{mccabe79}
{McCabe}, E.~M., {Connon Smith}, R., \& {Clegg}, R. E.~S. 1979, \nat, 281, 263

\bibitem[{{Papageorgiou} {et~al.}(1998){Papageorgiou}, {Demetropoulos}, \&
  {Lagaris}}]{merlin98}
{Papageorgiou}, D.~G., {Demetropoulos}, I.~N., \& {Lagaris}, I.~R. 1998,
  Comput. Phys. Commun., 109, 227

\bibitem[{{Press} {et~al.}(1986){Press}, {Teukolsky}, {Vetterling}, \&
  {Flannery}}]{numerical_recipes86}
{Press}, W.~H., {Teukolsky}, S.~A., {Vetterling}, W.~T., \& {Flannery}, B.~P.
  1986, Numerical Recipes (Cambridge: Cambridge University Press)

\bibitem[{{Russell}(1934)}]{russell34}
{Russell}, H.~N. 1934, \apj, 79, 281

\bibitem[{{Sauval} \& {Tatum}(1984)}]{sauvaltatum84}
{Sauval}, A.~J. \& {Tatum}, J.~B. 1984, \apjs, 56, 193

\bibitem[{{Shchukina} \& {Trujillo Bueno}(2001)}]{shchukina_trujillo01}
{Shchukina}, N. \& {Trujillo Bueno}, J. 2001, \apj, 550, 970

\bibitem[{{Socas-Navarro}(2003)}]{socas_navarro03}
{Socas-Navarro}, H. 2003, Neural Networks, 16, 355

\bibitem[{{Socas-Navarro}(2005)}]{socas_navarro05}
{Socas-Navarro}, H. 2005, \apj, 620, 517

\bibitem[{{Tejero Ordo\~nez} \& {Cernicharo}(1991)}]{tejero_ordonez91}
{Tejero Ordo\~nez}, J. \& {Cernicharo}, J. 1991, Modelos de Equilibrio
  Termodin\'amico Aplicados a Envolturas Circunestelares de Estrellas
  Evolucionadas (Madrid: IGN)

\bibitem[{{Tsuji}(1964)}]{tsu64}
{Tsuji}, T. 1964, Ann. Tokio. Astron. Obs., 2nd ser. 9, 1

\bibitem[{{Tsuji}(1971)}]{tsuji71}
{Tsuji}, T. 1971, Ann. Tokio Astron. Obs. 2nd ser., 9, 1

\bibitem[{{Tsuji}(1973)}]{tsu73}
{Tsuji}, T. 1973, \aap, 23, 411

\end{thebibliography}

\end{document}